%% file: CorrectNet_v2.tex
\documentclass[conference]{IEEEtran}
\IEEEoverridecommandlockouts

\usepackage{color}
\usepackage{graphicx}
\usepackage{amsmath}
	
\usepackage{soul}

\usepackage{subcaption}
\usepackage{caption}
\usepackage[utf8]{inputenc}

\makeatletter
\def\hlinewd#1{%
\noalign{\ifnum0=`}\fi\hrule \@height #1 %
\futurelet\reserved@a\@xhline}
\makeatother
\usepackage{booktabs}
\usepackage{multirow}

\usepackage[binary-units=true]{siunitx}

\usepackage[algoruled,noline,longend,linesnumbered]{algorithm2e}

\usepackage{pgfplots}
\usepackage{pgfplotstable}
\usepackage{siunitx}

\makeatletter

\usepackage{amsthm}

\usepackage[small,compact]{titlesec} 
\titlespacing{\section}{2pt}{2pt}{2pt}
\titlespacing{\subsection}{2pt}{2pt}{2pt}
\titlespacing{\subsubsection}{2pt}{2pt}{2pt}

\begin{document}

\graphicspath{{Fig/}}
\def\figname{Figure}
\def\algname{Algorithm}

\newcommand{\papertitle}{CorrectNet: Robustness Enhancement of Analog In-Memory Computing for Neural Networks
by Error Suppression and Compensation
}

\title{\papertitle}
\author{

}
\author{
\IEEEauthorblockN{Amro Eldebiky$^1$,
Grace Li Zhang$^2$,
Georg B\"ocherer$^3$,
Bing Li$^1$,
Ulf Schlichtmann$^1$}
\IEEEauthorblockA{$^1$Technical University of Munich, $^2$TU Darmstadt, $^3$Huawei Munich Research Center}
\IEEEauthorblockA{Email: \{amro.eldebiky, b.li, ulf.schlichtmann\}@tum.de, grace.zhang@tu-darmstadt.de, georg.bocherer@huawei.com}
}

\maketitle



\input{abstract}

\input{introduction}

\input{preliminaries}

\input{methodology}

\input{results}

\input{conclusion}

\input{ack}

\let\oldbibliography\thebibliography
\renewcommand{\thebibliography}[1]{%
\oldbibliography{#1}%
\fontsize{5.9pt}{5.9}\selectfont
\setlength{\itemsep}{0.2pt}%
}

\bibliographystyle{IEEEtran}
\bibliography{IEEEabrv,CONFabrv,bibfile}

\end{document}

%% file: abstract.tex
\begin{abstract}


The last decade has witnessed the breakthrough of deep neural networks (DNNs)
  in many fields.  With the increasing depth of DNNs, hundreds of millions of
  multiply-and-accumulate (MAC) operations need to be executed.  To accelerate
  such operations efficiently, analog in-memory computing platforms based on emerging
  devices, e.g., resistive RAM (RRAM), have been introduced. These acceleration
  platforms rely on analog properties of the devices and thus suffer from
  process variations and noise. Consequently, weights in neural networks 
  configured into these
  platforms can deviate from the expected values, which may lead to feature
  errors and a significant degradation of inference accuracy. To address this
  issue, in this paper, we propose a framework to enhance the robustness of
  neural networks under variations and noise.  First, a modified Lipschitz constant
  regularization is proposed during neural network training to suppress the
  amplification of errors propagated through network layers. 
Afterwards, error compensation is introduced at necessary locations determined by reinforcement learning 
  to rescue the feature maps with remaining errors. 
Experimental results
  demonstrate that inference accuracy of neural networks can be recovered
  from as low as 1.69\% under variations and noise back to more than 95\% 
  of their original accuracy,
  while the training and hardware cost are negligible.


\end{abstract}

%% file: introduction.tex
\section{Introduction} \label{sec:intro} 

Deep neural networks (DNNs) have been applied successfully in many fields,
e.g., image recognition \cite{krizhevsky2012imagenet} and language processing
\cite{chiu2018state}.  
DNNs achieve their accuracy using a large number of layers \cite{deeplearning}.  This
results in tens of millions weights and hundreds of millions of
multiply-and-accumulate (MAC) operations in a neural network. 
To accelerate these operations, 
analog in-memory computing
platforms based on emerging
technologies, e.g., resistive RAM (RRAM)  \cite{chi2016prime,shafiee2016isaac}, have been
introduced.  In such platforms, MAC operations are implemented by analog
devices based on Ohm's law and Kirchhoff's current law, so that a
high computation and energy efficiency can be achieved.

These analog-based computing platforms, 
however, 
suffer from manufacturing process variations and noise 
\cite{niu2010impact}. Accordingly, inference accuracy of neural networks
implemented with such platforms may degrade significantly in practice.  For
example, in an RRAM-based computing platform,  RRAM cells should be programmed to
specific conductances to represent weights of neural networks.  However,
variations of physical parameters of RRAM cells, e.g., cross-section area,
cause variability in their electrical properties.  Accordingly, when a programming voltage
is applied onto an RRAM cell,
the resulting conductance value 
under process variations and noise deviates from the
nominal value. 
Consequently, weights in neural networks 
may not be reflected accurately, 
and the feature maps at the output of layers can become
erroneous.  When such incorrect feature maps travel through subsequent layers
with deviated weights, the errors 
can be amplified, which
results in a significant degradation of inference accuracy and thus offsets the
advantages of these platforms in 
computation and energy efficiency \cite{liu2015vortex}. 
  

Several previous approaches have been proposed to tackle the accuracy
degradation problem due to hardware uncertainty.  
The method in \cite{charan2020accurate} applies
knowledge distillation to train a variation-aware model and replicates
some important weights into SRAM cells to enhance computational
robustness. Similarly, \cite{mohanty2017random} randomly selects 
some weights and maps them into on-chip memory for further training to improve accuracy. 
However,
these methods 
still require online retraining to restore the inference accuracy, which incurs extra
training and test cost.  
Trying to adjust weights to absorb the effect of variations,
\cite{liu2015vortex} proposes a variation-aware training approach  by
compensating the impact of device variations and including it in the loss
function.
The methods in \cite{9116244, long2019design} train neural networks statistically by directly modeling
variations as functions of random variables.  \cite{chen2017accelerator}
follows a similar approach regarding variation-aware mapping which
represents large weights by RRAM cells with lower variations in the crossbar
and trains the neural network adaptively online.  
Such approaches, however,
either need prior knowledge of the variation profile, 
which requires testing and measurement of each manufactured chip, or are limited
to neural networks with only a small depth.


Different from the approaches above, in this paper, we introduce a method to
mitigate the effects of weight variations and noise in analog in-memory computing platforms 
by two
techniques: error suppression and error compensation. 
The key contributions of this work are summarized as follows: 
\vspace{-3pt} 
\begin{itemize}
    \item Error suppression is realized by training neural networks under modified
    Lipschitz constant regularization,
    so that weight variations do not
    cause the amplification of errors resulting from previous layers. 
    \item 
      Error compensation is introduced to recover feature maps from potential
      errors.
      The locations of error
      compensation and the number of filters for compensation are determined by reinforcement learning to achieve a balance
      between accuracy recovery and computational cost.
    \item Experimental results demonstrate that the inference accuracy of
      neural networks can be recovered from as low as 1.69\% under variations and noise 
      back to more than 95\% of their original accuracy, while the training and hardware
      cost are negligible.
\end{itemize}

The rest of this paper is organized as follows. In
Section~\ref{sec:preliminaries}, we explain the background and motivation of
this work.  In Section~\ref{sec:methods}, we introduce the techniques of 
error suppression by Lipschitz constant regularization and error compensation
for sensitive layers.  Experimental results are reported in
Section~\ref{sec:results} and
conclusions are drawn in Section~\ref{sec:conclusion}.

%% file: preliminaries.tex
\section{Preliminaries and motivation}

\label{sec:preliminaries} 

\begin{figure}
    \centering
    \includegraphics{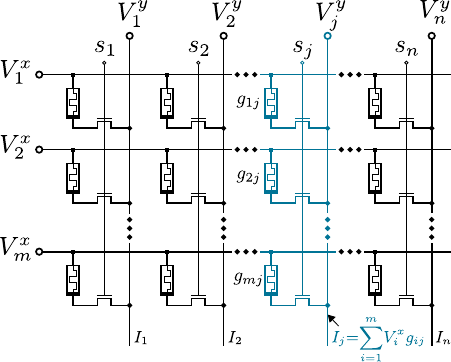}
    \caption{The structure of RRAM crossbar.}
    \label{fig:crossbar}
\end{figure}

To accelerate DNNs, 
emerging platforms with analog 
devices, e.g., RRAM, have a huge advantage in computation and energy efficiency. 
Figure~\ref{fig:crossbar} illustrates the 
structure of an RRAM-based crossbar, 
where   
RRAM cells sit 
at the crossing points and transistors are used to enable 
RRAM cells. 
To implement MAC operations, RRAM cells are first programmed to the target conductance 
values to represent weights of DNNs. 
Afterwards, voltages are applied on the horizontal wordlines while voltages on the 
vertical bitlines are connected to ground.    
The resulting current in an RRAM cell 
is thus the multiplication result of the voltage and its conductance. 
The accumulated currents at the bottom of each column is 
the addition result. 


Analog accelerators, however, are inherently susceptible to variations and noise from 
manufacturing process and operation environments, respectively.  
These variations and noise cause 
the programmed conductances to deviate from the target values. Accordingly,   
weights of neural networks represented by conductances of RRAM cells 
vary from their
nominal values, leading to a degradation of inference accuracy.  

To demonstrate the effect of variations and noise on the inference accuracy of neural networks, 
we use a log-normal distribution to inject variations 
into weights of neural networks as an example. This log-normal variation model is widely adopted, e.g., 
in \cite{long2019design,chen2017accelerator,liu2015vortex}. 
\begin{align}
  w&=w_{nominal}*e^{\theta} \label{eq1}\\
    \theta &\sim N(0, \sigma^2) \label{eq2}
\end{align}
where $w_{nominal}$ is the nominal value of a weight after training and
$\theta$ is a 
Gaussian random variable with $\sigma$ as its
standard deviation. 
For different weights, their corresponding variables $\theta$s are independent. 
 
\begin{figure}
    \centering
    \includegraphics[width=\linewidth]{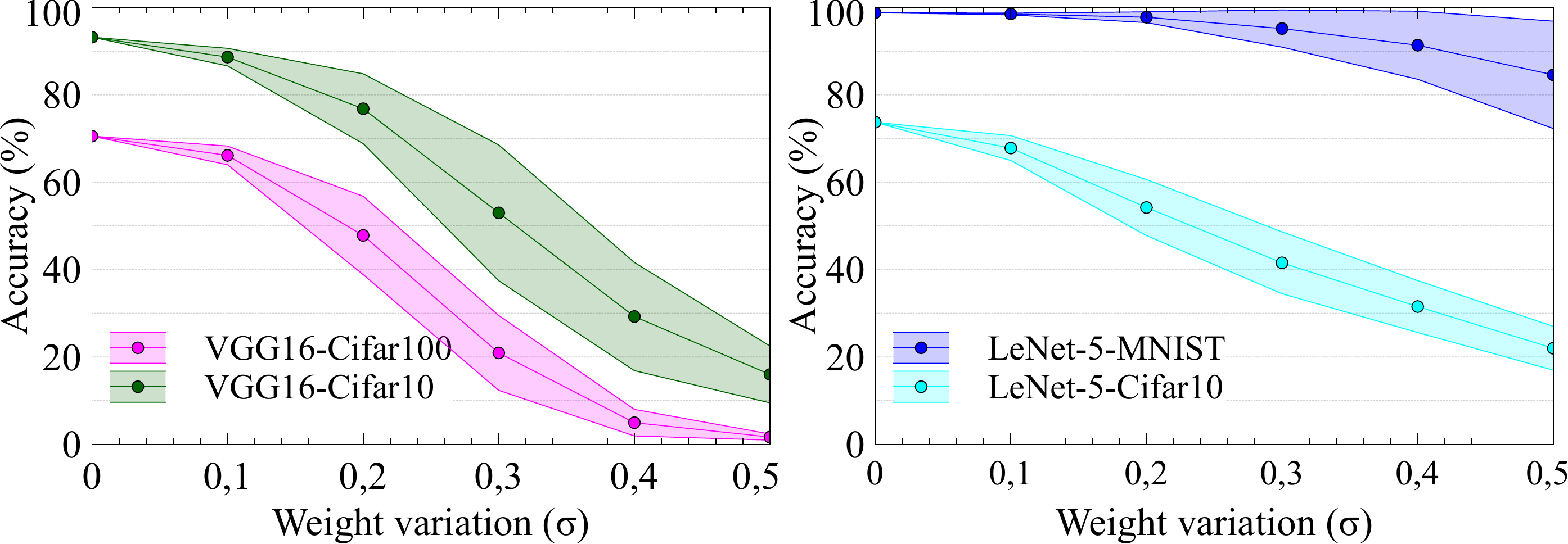}
    \caption{Inference accuracy degradation of neural networks under variations
    in weights.}
    \label{fig:1}
\end{figure}

Figure~\ref{fig:1} shows the mean values and the standard deviations of the
inference accuracy of VGG16 \cite{simonyan2014very} and LeNet-5
\cite{lecun1989backpropagation} on Cifar100, Cifar10, and MNIST datasets under
different levels of weight variations using the model in (\ref{eq1}) and (\ref{eq2}). 
The solid lines in the middle of the
ranges represent the mean values and the ranges represent the standard
deviations.  
According to Figure~\ref{fig:1}, even with relatively small variations, the
accuracies of the neural networks have degraded noticeably. As the amount of
variations increases, the accuracy drops even more drastically, which makes the neural
networks unusable in practice. 
In addition, VGG16 with more layers 
exhibited a more drastic accuracy degradation
than LeNet-5. The reason is that 
as these data propagate through more layers, not only are
further variations accumulated but the deviations in early layers can also
be amplified by the computation in later layers.

%% file: methodology.tex
\section{Design methodology for error suppression and compensation}
\label{sec:methods}

In this paper, we propose to suppress error propagation through layers by
applying a modified Lipschitz constant regularization \cite{gouk2021regularisation,cisse2017parseval,lin2019defensive}. 
To further enhance inference accuracy, 
an error compensation for selected layers is proposed. 
%
With these techniques, the inference accuracy of neural networks can be
recovered effectively 
to enable their execution on analog accelerators for energy-efficient computing. 
The proposed method is very general and can be applied into any analog in-memory computing platform for neural networks by 
adapting the variation model according to the corresponding analog devices.  


%
\subsection{Lipschitz constant regularization for error suppression}

A function $f: X \rightarrow Y$ is Lipschitz constrained 
\cite{cisse2017parseval,gouk2021regularisation} if it satisfies a certain 
p-distance metric
\begin{equation}
  \left| f(\mathbf {x_1})-f(\mathbf {x_2}) \right|_p \leq k \left| \mathbf {x_1 -
  x_2} \right|_p,\; \; \forall \mathbf{x_1,x_2} \in X
    \label{eq3}
\end{equation}
where the p-norm $\left| \cdot \right|_p$ calculates the p-distance metric
between two vectors.  For the function $f$, the smallest value of the
non-negative constant $k$ is denoted as the Lipschitz constant $L(f)=k$ and $f$
is said to be k-Lipschitz.  The Lipschitz constant $L(f)$ describes how $f$
scales with respect to its input. 
If $L(f)$ is larger than 1, any change in the input is amplified by $f$;
otherwise, the change is suppressed. For multiple functions $f_1, \dots f_l$
with Lipschitz constants $k_1, \cdots k_l$, their   
composition is also Lipschitz constrained as
\begin{align}
  &f = (f_l \circ f_{l-1}  \circ ... \circ f_1)(x)
    \label{eq4}\\
     &L(f) \leq k_{l} \cdot k_{l-1}\cdot ... \cdot k_1.
     \label{eq5}
\end{align}
In other words, the Lipschitz constant of the composition function is upper
bounded by $k_{l} \cdot k_{l-1}\cdot ... \cdot k_1$.

\begin{figure}
    \centering
    \includegraphics{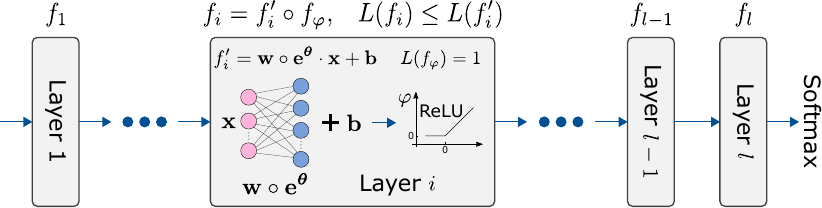}
    \caption{Lipschitz constant regularization and composition in a neural
    network.}
    \label{fig:composition}
\end{figure}

The composition of functions and the Lipschitz constant of the composition
can be used to bound
the forward propagation of errors in a neural network, because forward
propagation of a neural network can be considered as a composition of operations
of successive layers, as shown in Figure~\ref{fig:composition}. 
If the $i$th layer of a neural network realizes a
function $f_i(\mathbf{x})$, the function of the neural network with $l$ layers can thus
be written as (\ref{eq4}). 

\begin{figure}
    \centering
    \includegraphics[scale=0.9]{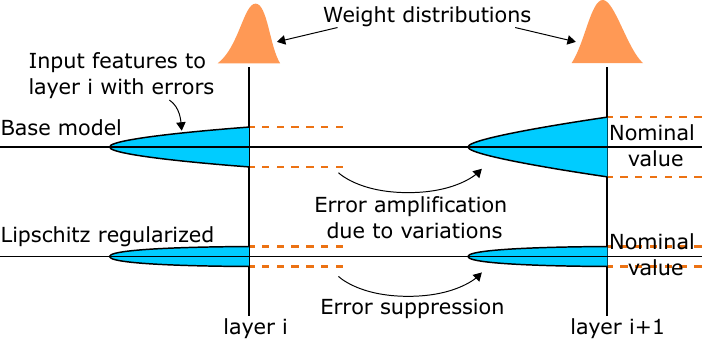}
    \caption{Lipschitz constant regularization for error suppression.}
    \label{fig:2}
\end{figure}

The concept of suppressing errors in neural networks can be illustrated in
Figure~\ref{fig:2}.  When specific data, e.g., an image, travels through the
neural network, the inputs to the $i$th layer may differ from the nominal
values, because the variations in the weights of the first $i-1$ layers cause
changes in their outputs and thus in the inputs of the $i$th layer.
The task of error suppression is thus to train the
neural network to obtain a set of weights that
limit the deviation of the outputs of layers of the neural network from
their nominal values when variations are considered. This training is
implemented based on the composition of the functions of the layers in the
neural network and Lipschitz constant regularization. 
 

Assume that the nominal inputs to the $i$th layer are written as $\mathbf{x_1}$
and the inputs affected by variations in the first $i-1$ layers are written as
$\mathbf{x_2}$, then the deviation of the outputs of the $i$th layer from the
nominal values can be evaluated as $\left| f_i(\mathbf{x_1}) -f_i(\mathbf{x_2})
\right|_p$, where $f_i(\cdot)$ is the function of the $i$th layer converting its
inputs to the outputs. To suppress this deviation, called \textit{error}
henceforth, we will use (\ref{eq3}) with $k\le 1$. In other words, errors will
not be amplified after a layer is traveled through.  According to the
composition in (\ref{eq4}), if all the layers in the neural network can meet
the Lipschitz constraint with $k\le 1$, the errors at the outputs of the neural
network will also be restrained according to (\ref{eq5}).

For a specific layer in the neural network, its function $f_i$ can be expressed
as the composition of $f_i^\prime$ and $f_\varphi$, as illustrated in
Figure~\ref{fig:composition}. $f_i^\prime= \mathbf{w\circ e^{\boldsymbol\theta}
\cdot \mathbf{x}}+\mathbf{b} $ implements the matrix-vector multiplication and
the sum with bias, in which $\mathbf{b}$ is the bias vector and $\mathbf{w}$
is the weight matrix of the layer.  $\mathbf{w\circ e^{\boldsymbol\theta}}$ is
the element-wise multiplication of the weight values with the random variables
$\mathbf{e^{\boldsymbol\theta}}$ to incorporate the effect of variations
according to (\ref{eq1}).  $f_\varphi$ is the ReLU function. The ReLU
function does not amplify any deviations and its Lipschitz constant is always 
equal to 1. Therefore,  
we only need to constrain the Lipschitz constant of $f_i^\prime$ to suppress
error amplification, as
\begin{align}
    \label{lip1}
      \left| (\mathbf{w\circ e^{\boldsymbol\theta} \cdot
     \mathbf{x_1}}+\mathbf{b}) - (\mathbf{w\circ e^{\boldsymbol\theta} \cdot
  \mathbf{x_2}}+\mathbf{b})\right|_p 
  &\leq \; k \left| \mathbf{x_1} - \mathbf{x_2} \right|_p\\
    \label{lip3}
    \Leftrightarrow \;\left| \mathbf{w\circ e^{\boldsymbol\theta}}
    \cdot (\mathbf{x_1}-\mathbf{x_2})\right|_p 
  &\leq \; k \left| \mathbf{x_1} - \mathbf{x_2} \right|_p\\
  \Leftrightarrow \frac{\left| \mathbf{w\circ e^{\boldsymbol\theta}}
    \cdot (\mathbf{x_1}-\mathbf{x_2})\right|_p}{\left| \mathbf{x_1} -
    \mathbf{x_2} \right|_p}
    &\leq \; k. \label{eq:lipcond}
\end{align}

According to the definition of the p-norm of a matrix $\parallel.\parallel_p$, 
the condition in (\ref{eq:lipcond}) can be expressed further as 
\begin{align}
    sup\left(\frac{ \left| \mathbf{w\circ e^{\boldsymbol\theta}}
    \cdot (\mathbf{x_1}-\mathbf{x_2})\right|_p}
  {\left| \mathbf{x_1} - \mathbf{x_2} \right|_p}\right) = 
  \parallel\mathbf{w\circ e^{\boldsymbol\theta}}\parallel _p\le k \label{lip4}
\end{align}
which shows that error propagation in a layer can be suppressed by constraining
$\parallel \mathbf{w\cdot e^{\boldsymbol\theta}}\parallel_p$.



%
%

Since $\boldsymbol{e^\theta}$ in (\ref{lip4}) represents a matrix of
independent random variations, $\parallel\mathbf{w\circ
e^{\boldsymbol\theta}}\parallel_p$ cannot be evaluated directly. To address this
problem, we use ${\mu_{e^\theta} + 3\cdot \sigma_{e^\theta}}$ to bound the random
variable ${e^\theta}$.
Since ${e^\theta}$ has a lognormal distribution, ${\mu_{e^\theta} + 3\cdot
\sigma_{e^\theta}}=e^{\frac{\sigma^2}{2}} + 3
\sqrt{(e^{\sigma^2}-1)e^{\sigma^2}}$, in which $\sigma$ is the standard
deviation of $\theta$. Accordingly, (\ref{lip4}) can be converted into
\begin{equation}
  \parallel\mathbf{w}\parallel_p \le  \lambda,
\; \; \lambda=  \frac{k}{e^{\frac{\sigma^2}{2}} + 3
  \sqrt{(e^{\sigma^2}-1)e^{\sigma^2}}}.
    \label{eq8}
\end{equation}

In the proposed method, we use the $L^2$ norm to bound $\mathbf{w}$ in
(\ref{eq8}), which corresponds to the spectral norm of $\mathbf{w}$. The spectral norm of a matrix is the maximum singular value of the matrix. To 
limit the spectral norm of the weight matrix, 
a regularization term is added to the loss function when training the neural
network,
as
\begin{equation}
  Loss = L_{ce} + \beta * \sum_{\mathbf{w_i} \in \mathbf{W}} \parallel
  \mathbf{w_i^Tw_i} - \lambda^2 \mathbf{I}\parallel^2
    \label{eqloss}
\end{equation}
where $L_{ce}$ is the original cross-entropy loss and $\mathbf{w_i}$ is the weight
matrix of the $i$th layer, $\mathbf{W}$ is the set of weight matrices of all layers,
and $\beta$ is a regularization hyperparameter. The added
regularization term keeps the weight matrix orthogonal to limit its maximum
singular value by $\lambda$ and hence limit its spectral norm.


The extra regularization term in (\ref{eqloss}) is calculated 
from all the layers in the neural network.
In applying (\ref{eqloss}), $\lambda$ is determined by setting the Lipschitz
constant $k$ to 1, so that errors will not be amplified. 
According to (\ref{eq4}) and (\ref{eq5}), this composition can thus suppress
error propagation in the whole neural network.

\subsection{Error compensation for accuracy recovery}

To further enhance inference accuracy, 
we propose to introduce light-weighted error
compensation to the early layers to recover the inference accuracy.
This error compensation incurs only a marginal  
computational cost, so that it can be executed on digital circuits 
\cite{hybrid2022} and 
is thus considered immune from
the effect of variations.

\begin{figure}
    \centering
    \includegraphics{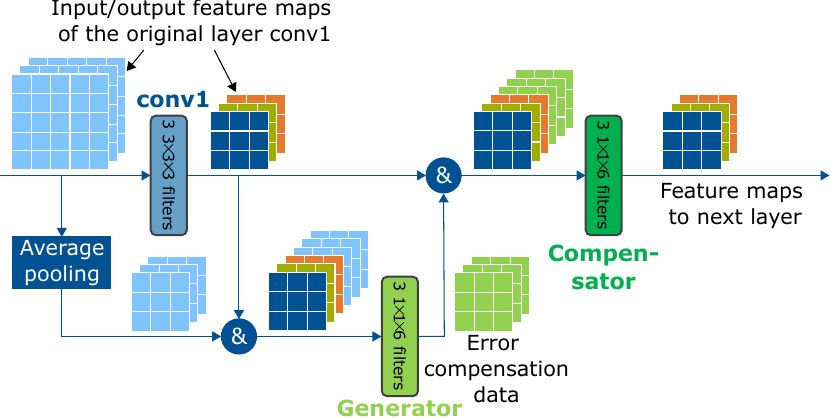}
    \caption{Error compensation for a convolutional layer.} 
    \label{fig_correction}
\end{figure}

Inspired by the concept of error correction in communication systems, which 
has been proposed as early as
in
\cite{shannon1948mathematical}, 
we generate error compensation data from the input and the output of 
a layer.
    The error compensation data are then used by a compensator to 
reduce the errors propagated through this layer. 
%
The concept of applying error compensation to a convolutional layer in a neural
network is illustrated in Figure~\ref{fig_correction}.
The generator is a small convolutional layer. The input and output feature
maps of the original convolutional layer (conv1) are concatenated and used as
the input of the generator to produce the compensation data.
Since the dimensions
of input feature maps and output features maps of the original layer do not
match, we apply average pooling to reduce the dimension of the input feature
maps so that they can be concatenated with the output feature maps and
processed
by the same filter.
%
%
%

The generator 
contains $m$ $1\times 1 \times (l+n)$ filters, where $l$ and $n$ are the number of 
input feature maps and the number of output feature maps in the original layer (conv1), respectively. 
In the example in Figure~\ref{fig_correction}, we use $l=n=3$ to explain the working mechanism of the generator. 
We use 1$\times$1 kernel dimension for two advantages. First, its computational
overhead is low. Second, the generated compensation data has the same dimension
of the output feature maps of the original layer, so that error compensation can
also be implemented with simple $1\times 1$ kernels in the compensator. 
The number of filters $m$ indicates the number of output feature maps produced by the generator, e.g., 3 in Figure~\ref{fig_correction}. 
The larger $m$ is, the larger is the computational cost and the more robust the 
neural network potentially becomes. 

The compensator is also a convolutional layer taking the compensation data
generated by the generator and the output feature maps of the original layer as
input. This compensator contains $n$ $1\times 1 \times (n+m)$ filters. 
The $n$ filters in the compensator guarantee that the compensator produces the same number of 
feature maps as the original layer. 
The $n+m$ kernels are required due to 
the concatenation of the output feature maps of the original layer and the outputs of the generator.  


When training the weights in the generators and compensators introduced to some
layers in a neural network, the weights in the original layers are fixed to the
values after applying Lipschitz constant regularization and stay non-trainable,
while the weights in the generators and compensators are kept trainable.  The
generators and compensators are then trained with the same training data using
the original cost function.
In this training, variations are sampled
statistically and applied to the corresponding weight values in the original
layer during each training batch. The weights in the generators and
compensators
are then adjusted in backward propagation to reduce the cost function.  





\begin{figure}
    \centering
       \includegraphics{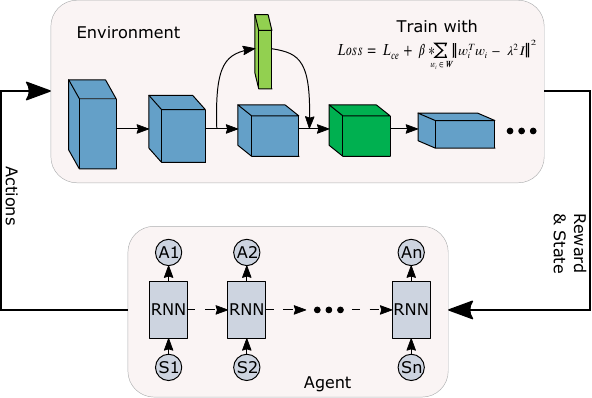}
    \caption{RL search for locations and filter numbers of error compensation. 
$A_1$, $A_2$, ... $A_n$ is the action sequence generated by the policy neural network in the agent. $S_1$, $S_2$, ... $S_n$ is  
the sequence to represent the state of the environment.} 
    \label{fig:reinforcement}
\end{figure}

To determine the locations of error compensation, 
we first inject variations into the layers from the last one backwards to the $i$th
layer. 
When $i$ is reduced, more layers contain variations, leading to a decreased
inference accuracy. The candidates of the neural network layers for error
compensation are then determined as the first $i$ layers when the variations in
the $i$th layer to the last layer lead to an inference accuracy lower than 95\%
of the original accuracy.
In the next step, we will apply RL to select concrete layers in the first $i$ layers 
and their numbers of filters for error compensation. During this process all the layers of a neural network 
are injected variations. 
Figure~\ref{fig:reinforcement} illustrates the application of RL, 
where the environment is defined as 
the neural network trained with error suppression and compensation whose locations and the filter numbers are determined by RL.
The state of the environment is the specified locations of error compensation 
and the corresponding number of filters. 
To represent the state, 
we use a sequence of $n$ floating point numbers, e.g., $S_1$, ..., $S_i$, ... $S_ {n}$, where 
$S_i$ 
is the ratio of the number of filters in the generator to the number of filters in the original $i$th layer.  
For example, $S_2=0.5$ means that 
the number of filters in the corresponding generator of the second layer is 
0.5 times the number of filters of the original second layer. 
$S \leq 0$  means no insertion of error compensation at a layer. 
To generate such a sequence for the environment state, we adopt recurrent neural network as the policy neural network in the agent. 

To train 
the policy neural network, 
we define a reward function as follows 
\begin{equation}
R= 
\begin{cases}
    acc_{avg} - acc_{std}-overhead,& \text{if } overhead\leq limit\\
    -overhead,              & \text{otherwise}
\end{cases} \label{reward}
\end{equation} 
where 
$acc_{avg}$, and $acc_{std}$ are the average and the standard deviations of the
inference accuracy of the trained neural network under the current environment state, respectively. 
The \textit{overhead} represents the ratio of the number of weights in the compensation
layers to the 
number of weights in the original neural network. 
To avoid large computational cost incurred by error compensation during the search process, a maximum number of weights 
for error compensation is set. 
If the \textit{overhead} of a solution exceeds the maximum \textit{limit} in (\ref{reward}), 
a negative reward is generated directly, 
so that the training of neural networks with error suppression and compensation 
in the current iteration 
can be skipped to make the agent learn fast and thus reduce execution time.  
In the experiments, 1\%, 2\%, and 3\% weight overhead were used as the maximum limit and 
the solution that generates 
the best accuracy was selected as the result. 
To determine the parameters of the policy neural network, 
a given number of episodes, each of which includes a specific number of learning iterations, are used. 

%% file: results.tex
\section{Experimental results}\label{sec:results}

To evaluate the proposed framework, 
two neural networks, VGG16 \cite{simonyan2014very} and LeNet-5
\cite{lecun1989backpropagation} were tested against three different datasets,
Cifar100, Cifar10 
and MNIST.
The neural networks were trained with Nvidia Quadro RTX 6000 GPUs.
The variation model used in the experiments 
was the log-normal distribution of weights mapped onto RRAM cells (\ref{eq1})
and (\ref{eq2}) as in \cite{long2019design,chen2017accelerator,liu2015vortex}. 
In (\ref{eq8}) and (\ref{eqloss}), $k$ is
set to 1 to suppress the propagation of errors and $\lambda$ is determined
based on the variations and the value of $k$.
In the experiments, the network weights were sampled 250 times according to
the variation model in (\ref{eq1}) and (\ref{eq2})
and inference accuracy was evaluated for each sample.  



Table~\ref{tab:results} summarizes the results showing the performance of the
CorrectNet framework when $\sigma$ in (\ref{eq1}) and (\ref{eq2}) was set to
0.5. This variation setting is already very large for variations in RRAM cells
\cite{long2019design,chen2017accelerator,liu2015vortex}.  The column $\sigma=0$
in Table~\ref{tab:results} shows the inference accuracy of the original neural
networks without variations. When variations of the amount $\sigma=0.5$ were
applied to the weights in the original neural networks, the inference accuracy
degraded significantly down to as low as 1.69\% on average for VGG16-Cifar100. With the
CorrectNet framework, this accuracy can be recovered back to 67.01\% on average, more than
95\% of the inference accuracy without variations. In Table~\ref{tab:results},
the lowest ratio of the inference accuracy of CorrectNet to the original inference accuracy without variation is 
92\% (LeNet-5-Cifar10, 74.9\%/80.89\%=92.6\%).

In the CorrectNet framework, the Lipschitz training method does not require
extra resource or incur extra computational cost. The overhead results from the
compensation layers.  In Table~\ref{tab:results}, the weight overhead 
is calculated as the percentage of the number of weights in
the compensation layers to the number of weights in the original neural network.
Compared with the computational operations 
in the original neural networks, the weight overhead of CorrectNet is
marginal while an effective accuracy recovery is still achieved. 
The numbers of
compensation layers in the neural networks after applying CorrectNet
are also shown in the last column of Table~\ref{tab:results},
which confirm that only some layers in the original neural networks
require error compensation after error suppression with Lipschitz constant
regularization is applied. 
\setlength{\tabcolsep}{2.5pt}
\begin{table}

  \footnotesize
    \centering
    \caption{Experimental results of CorrectNet.}
    \vspace{-5pt}
    \begin{tabular}{ccrrcrcrc}
        \hline
         \multirow{3}{*}{Network-Dataset} &  &\multicolumn{4}{c}{Inference accuracy} & &
           \multicolumn{2}{c}{CorrectNet} \\
         \cline{3-6}
          &  &\multicolumn{2}{c}{Original network} & & CorrectNet&&\multicolumn{2}{c}{overhead} \\
         \cline{3-4}\cline{6-6} \cline{8-9}
          & & \multicolumn{1}{c}{$\sigma=0$} & \multicolumn{1}{c}{$\sigma=0.5$}
          && \multicolumn{1}{c}{$\sigma=0.5$}& &\multicolumn{1}{c}{Weights}& \#Layers \\
         \hline
         VGG16-Cifar100  && 70.52\% & 1.69\%  && 67.01\% && 1.03\%  & 4\\
         VGG16-Cifar10   && 93.2\%  & 16.01\% && 91.29\% && 0.58\% & 3\\
         LeNet-5-Cifar10 && 80.89\% & 25.29\% && 74.9\%  && 3.47\% & 1\\
         LeNet-5-MNIST   && 98.79\% & 84.58\% && 97.47\% && 5\%    & 2\\
         \hline
    \end{tabular}
    \label{tab:results}
\end{table}

\begin{figure}
    \centering
    \includegraphics[width=\linewidth]{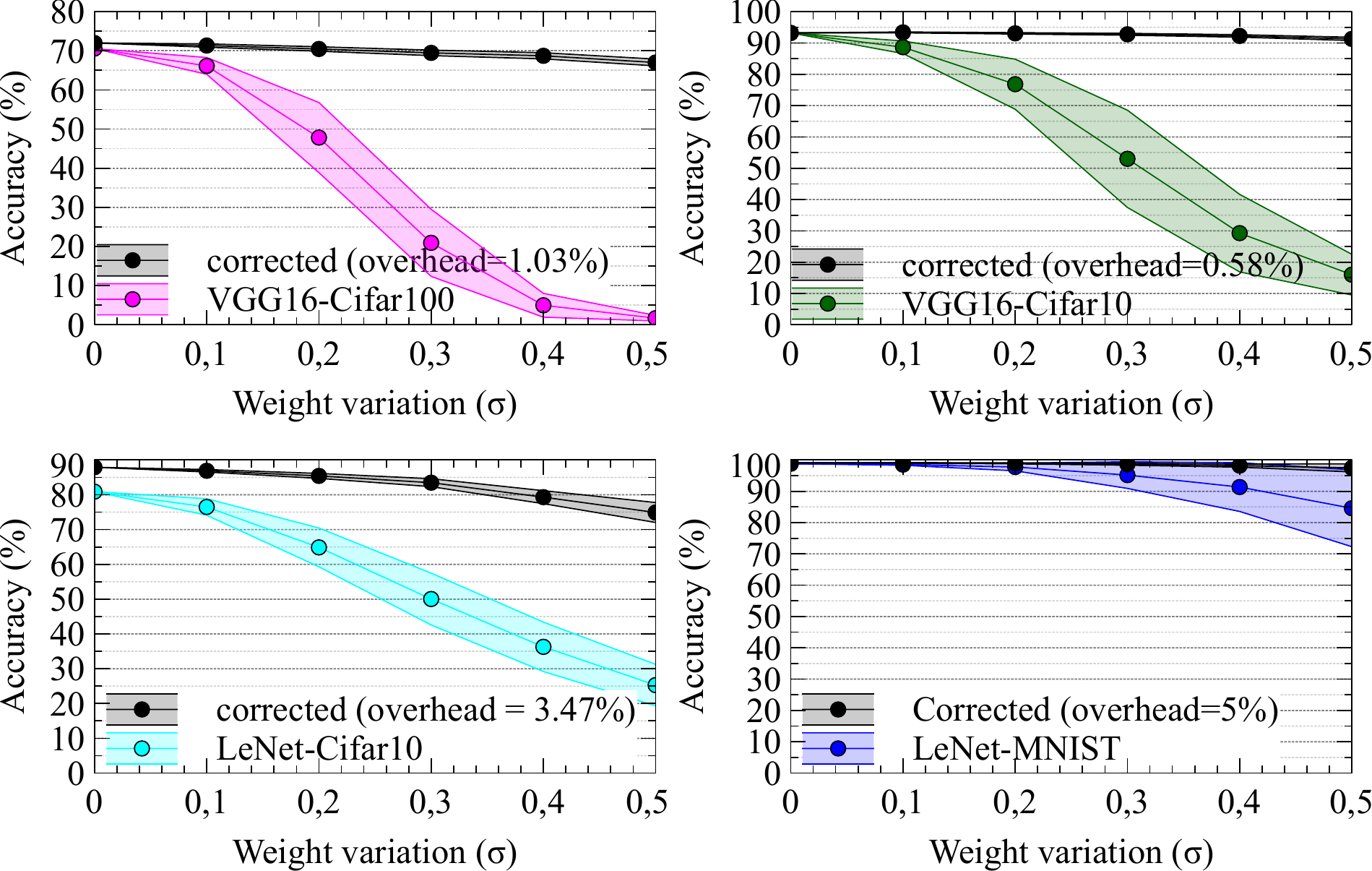}
    \caption{Accuracy of CorrectNet under different variations.}
    \label{fig:best}
\end{figure}

To demonstrate the capability of CorrectNet under different variation
scenarios, we tested this framework using the same combinations of neural
networks and datasets under different variation settings. 
The results are shown in Figure~\ref{fig:best}. In
each of these figures, we compare the inference accuracy of CorrectNet and the
original neural network. The mean values are shown with the solid lines while
the ranges show the standard deviations. In all these test cases, CorrectNet
has demonstrated an effective and robust trend to recover inference accuracy
under different variations.

\begin{figure}[t]
    \centering
    \includegraphics[width=0.95\linewidth]{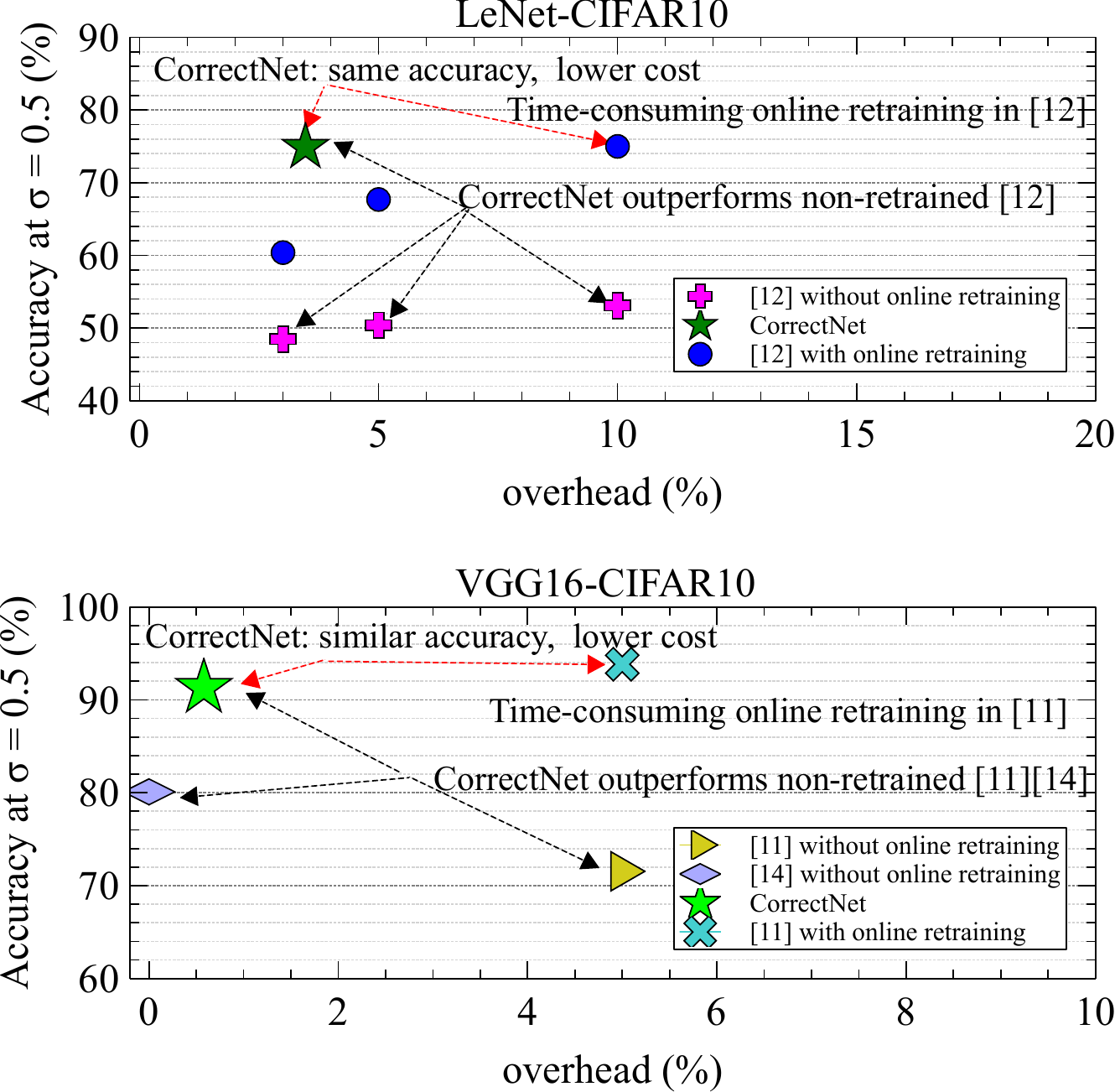}
    \caption{CorrectNet versus the state of the art \cite{mohanty2017random,charan2020accurate,long2019design}. }
    \label{fig:compare}
\end{figure}

To evaluate CorrectNet, we also compare its results with those from  
\cite{mohanty2017random,charan2020accurate,long2019design} as shown in Figure~\ref{fig:compare}. 
The x-axis represents the overhead incurred by weights for error compensation, 
and the y-axis is the mean value of 
the inference accuracy under variations of $\sigma= 0.5$. 
According to this comparison, 
CorrectNet achieves a higher accuracy than \cite{charan2020accurate} and \cite{mohanty2017random} 
with a smaller overhead in case of non-online retraining. 
In the case with time-consuming online retraining in \cite{charan2020accurate} and 
\cite{mohanty2017random}, 
CorrectNet 
can also achieve a similar accuracy with a lower overhead while time-consuming online retraining is not needed. 
CorrectNet also outperforms \cite{long2019design} in accuracy with a slightly larger overhead.

\begin{figure}
    \centering
    \includegraphics[width=0.8\linewidth]{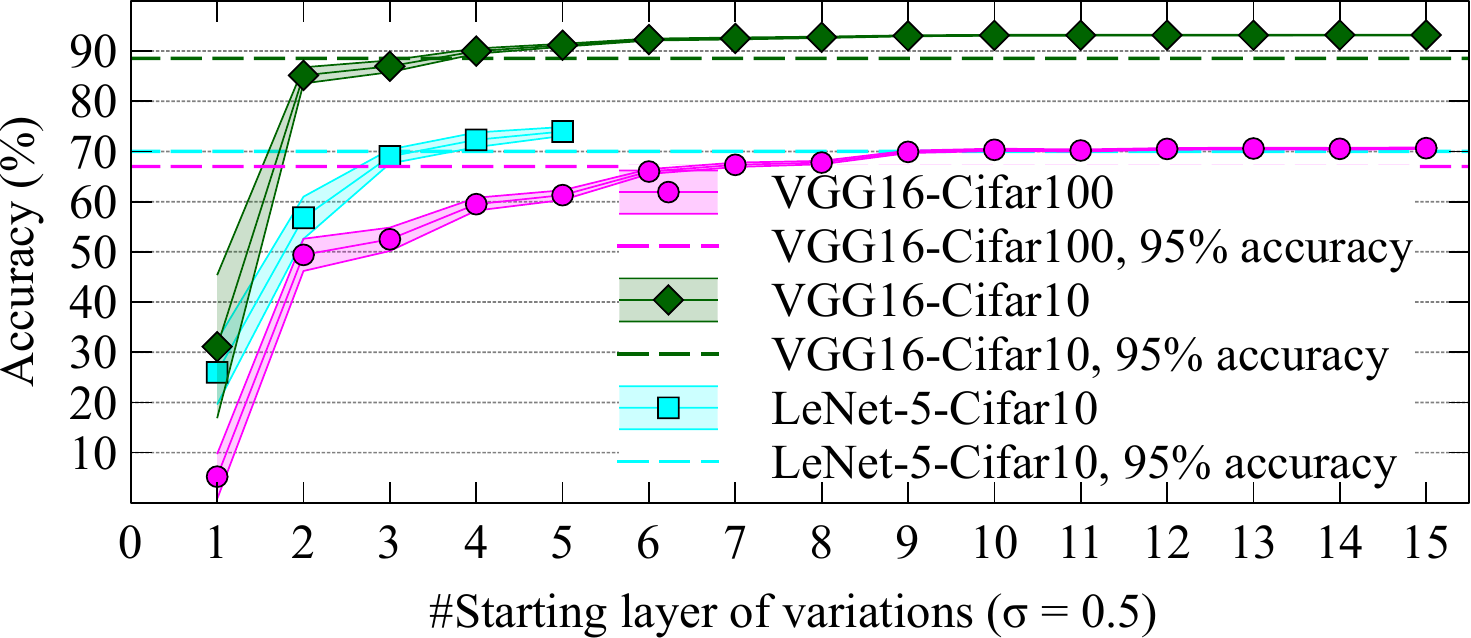}
    \caption{Lipschitz constant regularization against variations from
    a given layer to the last layer. }
    \label{fig:lip}
\end{figure}

To demonstrate the effectiveness of Lipschitz constant regularization, we added
variations to the weights from the $i$th layer to the last layer of the neural
networks after training with this regularization, while error compensation was
disabled.  Figure~\ref{fig:lip} shows the inference accuracy of the neural
networks with these variations from the $i$th layer to the last layer while 
$\sigma$ was set to
0.5.  The results corresponding to starting layer 1 on the x-axis are the cases
applying Lipschitz constant regularization to the whole neural networks without
error compensation.  From this figure, it can be observed that Lipschitz constant
regularization can counter variations in the late layers of the neural networks
effectively. 
But the inference accuracy of the neural networks is very sensitive to
variations in early layers and
the accuracy 
can only be recovered by error compensation in early layers to
achieve the results shown in Table~\ref{tab:results}.

In CorrectNet, the locations and parameters of error compensation are determined by RL. 
According to
Figure~\ref{fig:lip}, the first six layers of VGG16 processing the dataset
Cifar100 were selected as candidates to be evaluated in RL search.
Figure~\ref{fig:ga} shows the quality of the explored solutions for error
compensation with $\sigma$ set to 0.5. The x-axis shows the
weight overhead of compensation layers and the y-axis shows the corresponding
inference accuracy.  The range for each dot represents the standard deviation
of the inference accuracy. If all these six layers contain error compensation,
the overhead is 4.29\% while the mean value and the standard deviation of the
inference accuracy are 67.14\% and 0.83\%, respectively. In contrast, RL
determines that only four layers need error compensation and the mean value and
the standard deviation of the inference accuracy can be recovered to 67.01\%
and 0.87\%, respectively. 
This inference accuracy, 
which already reaches 95\% of the original inference accuracy, 
is comparable with that achieved by exhaustive error compensation  
where all six layers contain error compensation. 

%

%% file: conclusion.tex
\section{Conclusion}
\label{sec:conclusion}

In this paper, we have proposed the CorrectNet framework 
to recover inference accuracy of in-memory analog computing platforms under variations. 
The proposed framework consists of error suppression by
training with Lipschitz constant regularization 
and error compensation for sensitive layers.
With only a marginal overhead, the CorrectNet framework can recover
inference accuracy from as low as 1.69\% under variations and noise back to
more than 95\% of their original accuracy. 

\begin{figure}
    \centering
    \includegraphics[width=0.9\linewidth]{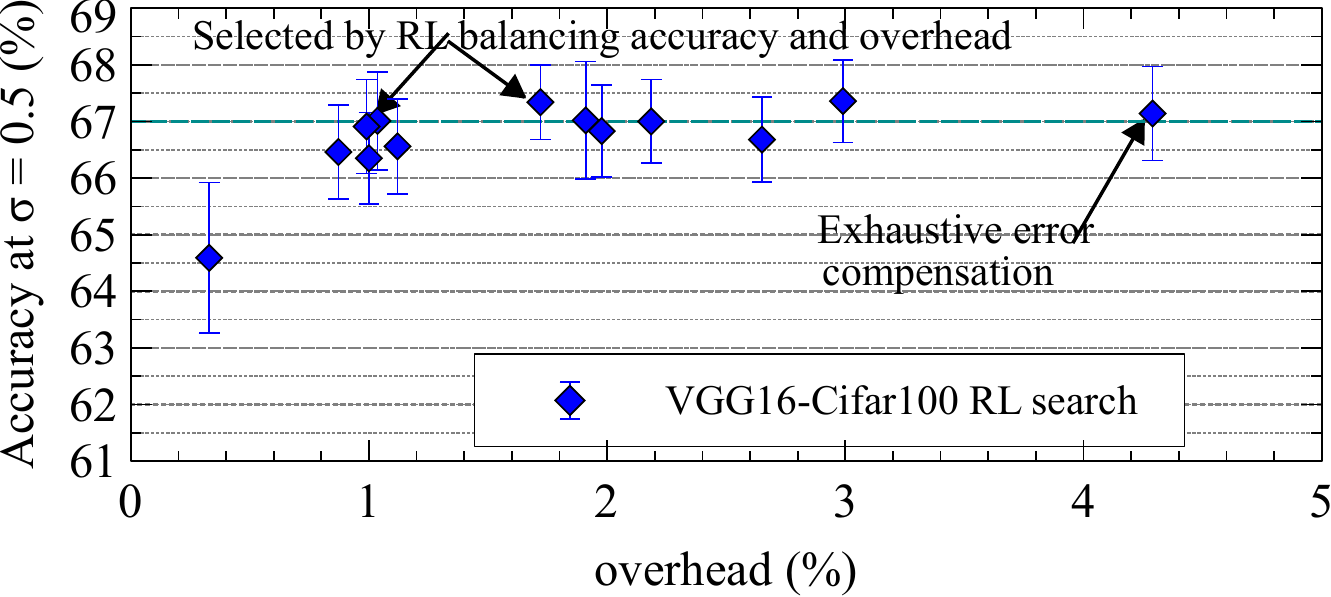}
    \caption{RL search for locations and parameters of error compensation.}
    \label{fig:ga}
\end{figure}

%% file: ack.tex
\section*{Acknowledgement}
This work is funded by the Deutsche Forschungsgemeinschaft (DFG, German Research Foundation) – 457473137.